\documentstyle[preprint,eqsecnum,aps,epsf,psfig]{revtex}
\newif\iftightenlines\tightenlinesfalse
\tightenlines\tightenlinestrue
\begin{document}

\title{Propagation of Muons and Taus at High Energies}
\author{S. Iyer Dutta$^1$, M. H. Reno$^2$, I. Sarcevic$^1$ and D. Seckel$^3$}
\address{
$^1$Department of Physics, University of Arizona, Tucson, Arizona 85721\\
$^2$Department of Physics and Astronomy, University of Iowa, Iowa City,
Iowa 52242\\
$^3$Bartol Research Institute, University of Delaware, Newark, Delaware 
19716}

\maketitle

\begin{abstract}

The photonuclear contribution to charged lepton energy loss 
has been re-evaluated taking into account HERA results on real and 
virtual photon interactions with nucleons. With large $Q^2$ processes 
incorporated, 
the average muon range in rock for 
muon energies of $10^9$ GeV is reduced by only 5\%\ compared with the standard
treatment. 
We have calculated the tau energy loss for energies up to
$10^9$ GeV taking into consideration the decay of tau.
A Monte Carlo evaluation of tau survival probability and range show that at
energies below $10^7-10^8$ GeV, depending on the material,
only tau decays are important. At higher energies 
the tau energy losses are significant, reducing the survival 
probability of the tau. We show that 
the average range for tau is shorter than its
decay length and reduces to  17 km in water for an incident tau energy
of $10^9$ GeV, as
compared with its decay length of 49 km at that energy. In iron, the average
tau range is 4.7 km for the same incident energy.

\end{abstract}

\section{INTRODUCTION}

Neutrino telescopes have the potential for detecting distant
sources of high energy neutrinos, for example, from Active Galactic Nuclei and 
Gamma Ray Bursters \cite{halzen}. Upward-going muons from muon neutrino conversions to
muons via charged current interactions with nuclei are the main signal
of muon neutrinos \cite{gqrs,gqrs2}. 
In addition, muons as well as neutrinos are produced 
in the atmosphere by cosmic ray interactions with air nuclei. 
Underground detector measurements of muon fluxes as a 
function of zenith angle are one way to determine the atmospheric muon flux as a function
of energy. At high energies, one expects the muon flux to reflect the onset of contributions 
from the production of charm in the atmosphere \cite{prompt}. 
A good understanding of the muon energy loss at high energies is an essential 
ingredient to the analysis of the high energy atmospheric muon flux. 

Recent SuperK measurements of atmospheric neutrinos strongly suggest 
${\nu_{\mu}}\rightarrow {\nu_{\tau}}$ oscillation \cite{sterile}
with large mixing angle of
$\sin^2{2\theta}>0.84$ and a neutrino mass squared difference of
$2\times 10^{-3}$ eV$^2<\Delta m^2< 6\times 10^{-3}$ eV$^2$ \cite{superk}.
Assuming ${\nu_\mu}\rightarrow {\nu_\tau}$ oscillation of extragalactic 
neutrinos with SuperK parameters, about half of the muon neutrinos 
are converted to tau neutrinos on the way to the Earth.
This leads to comparable fluxes of ultrahigh energy tau neutrinos and 
muon neutrinos at the Earth. Tau energy loss affects tau neutrino propagation
in the Earth, where tau neutrinos interact with nucleons to produce taus
which subsequently decay \cite{drs}. An understanding of tau energy loss
at very high energies could help with the interpretation of long
tracks produced by charged particles in large underground detectors.  

Future neutrino telescopes such as AMANDA, NESTOR, ANTARES and 
ICECUBE are aimed at detecting 
high energy events from extragalactic neutrino sources. 
The high energy 
behavior of muon and tau interactions with water or rock nuclei
have implications for event rates and the eventual unfolding of 
their respective parent neutrino flux. 

One way to characterize the energy loss of charged leptons is to consider the
average lepton energy loss per distance traveled ($X$ in
units of g/cm$^2$), which
can be expressed in the form
\begin{equation}
-\langle{dE\over dX}\rangle = \alpha +\beta E
\end{equation}
where $E$ is the lepton energy, $\alpha$ is 
nearly constant, determined
by the ionization energy loss,
and $\beta=\sum_i\beta^i$ is
weakly dependent on energy and due to radiative energy loss through
bremsstrahlung, pair production and photonuclear scattering. 
Generically,
\begin{equation}
\beta^i (E) = {N\over A}\int_{y_{min}}^{y_{max}} 
dy\  y{d\sigma^i(y,E)\over dy}
\end{equation}
where $y$ is the fraction of lepton energy loss in the radiative interaction:
\begin{equation}
y={E-E'\over E}
\end{equation}
for final lepton energy $E'$. The superscript $i$ denotes
bremsstrahlung (brem), pair production (pair) and photonuclear (nuc)
processes. Avogadro's number is $N$ and the atomic mass number of the target
nucleus is $A$. For rock, typical values for initial muons of energy
$\sim 100$ GeV are $\alpha\simeq 2.4$ MeV/(g cm$^{-2}$) and 
$\beta\simeq 3\times 10^{-6}$/(g cm$^{-2}$) \cite{lkv}.

At low energies, continuous energy loss by ionization dominates muon
propagation, but at higher energies (above $\sim 10^3$ GeV), losses
through pair production, bremsstrahlung and photonuclear interactions
dominate. In the case of the muon, pair production is the most important
mechanism, but for taus, the photonuclear process is at least as important
as pair production. The high energy extrapolation of the photonuclear
cross section has the largest theoretical 
uncertainty in the contributions to energy loss.

The experimental interest in high energy muons and taus
make it useful to reanalyze the
high energy photonuclear differential cross section. 
Bottai and Perrone\cite{bottai}
have reevaluated the photonuclear contributions using the HERA results for 
$\sigma(\gamma N)$.
In this paper, we evaluate the photonuclear differential cross section
using HERA results for real {\it and} virtual photon-nucleon scattering.
Here, we use a parameterization of the measured electromagnetic 
structure function $F_2$ which agrees well with data over the
full range of $Q^2$: from real photon interactions with nucleons to 
highly virtual photon interactions with nucleons.
One expects that the inclusion
of the $Q^2\gg0$ part of the photonuclear cross section will contribute
to larger muon energy losses at high energy. Indeed, this is what we
find at the highest muon energies considered. We use the same
expression for photonuclear interactions of taus with the correction
for the different lepton mass.

In the next section and Appendix A, 
we outline the formulas for evaluating
charged lepton energy loss, and we review the standard treatment
of the photonuclear cross section discussed by Bezrukov and
Bugaev in Ref. \cite{bb}. We outline the Monte Carlo program for
energy loss (and decay, for the tau).
In the following section, with details
in Appendix B, we describe our treatment of 
the photonuclear cross section using $F_2$.
In Section IV, we evaluate and compare our energy losses with
the standard treatment in the case of muons.
Section V includes  $\beta^i$, range and survival probability for taus
taking into consideration the decay probability.
In Section VI, we summarize our results and discuss some implications
of the new high energy photonuclear contributions to charged lepton
energy loss.

\section{Charged Lepton Energy Loss}

A standard compilation of the energy loss parameters
for muon energy loss was performed by Lohmann, Kopp
and Voss (LKV) \cite{lkv}. The standard formulas are reproduced
in Appendix A.
In evaluating the energy loss due
to ionization, the Bethe-Bloch formula \cite{bloch}
is used with parameters
listed in Table I {\cite{sternheimer}}.
Bremsstrahlung energy losses are 
computed via the differential cross section of Petrukhin and Shestakov
\cite{ps}. Improvements of the bremsstrahlung calculation \cite{newbrem} 
affect the muon 
intensity-depth relationship by only a few percent \cite{music}  
and should affect the tau energy loss even less.
The pair production differential cross section is parameterized by
Kokoulin and Petrukhin \cite{kp}. 

The standard  photonuclear differential 
distribution used in muon energy loss calculations is by
Bezrukov and Bugaev (BB) \cite{bb}.
Their differential cross section is based on a generalized vector dominance
model with off-diagonal contributions, and the nuclear shadowing is evaluated
using an optical model. 
The BB parameterization of the photonuclear
cross section is a function of $A^{1/3}\sigma_{\gamma N}(E)$, as
exhibited in Eqs. (A10) and (A11).
Our alternative to the BB differential cross section is described
in detail in the next section.

For quantitative comparisons of the two 
approaches to photonuclear energy loss, we use a one-dimensional 
Monte Carlo propagation program to evaluate survival
probabilities of muons and taus as a function of energy and depth $X$.
By using the $dE/dX$ formula and integrating, one can find the
``range of average energy loss'' $R_{\langle \Delta E\rangle}$.  
Lipari and Stanev (LS) 
have shown that this is different than the average range
$\langle R(E)\rangle$ \cite{ls}. 
With a stochastic treatment of bremsstrahlung, pair production and 
photonuclear interactions, fluctuations have the
effect of decreasing the average range relative to the range of average
energy loss. 
Accounting for fluctuations
has implications for the downward muon rates where the tails of the
survival probabilities are important \cite{ls}.
The Lipari-Stanev evaluation of the muon range 
relies on the Bezrukov-Bugaev photonuclear contribution.

The one-dimensional approximation, where the outgoing charged lepton 
travels
in the same direction as the incident lepton, should be
adequate for the high energies considered here.
More elaborate (three-dimensional) evaluations
have been performed by other authors, for example, in the
program MUSIC by Antonioli {\it et al.}\cite{music}.
The MUSIC program compares well with the one-dimensional 
Lipari-Stanev survival 
probabilities, 
both of which use the differential cross sections outlined in LKV. 

In constructing our Monte Carlo program,
we follow the standard procedure of splitting the radiative energy loss into
two terms, a continuous ``soft'' term for $y\leq y_{cut}$, and
a stochastic ``hard'' term for $y_{cut}<y\leq 1$ in
\begin{equation}
-{dE\over dX}=\alpha + {N\over A}E\int_0^{y_{cut}}dy\, y{d\sigma\over dy}
+ {N\over A}E\int^1_{y_{cut}}dy\, y{d\sigma\over dy}\ ,
\end{equation}
while the ionization is treated continuously. We have chosen
$y_{cut}=10^{-3}$ following Ref. \cite{music}. For taus, we have 
added a stochastic term for tau decay.

\section{Photonuclear Cross Section using $F_2$}

The photonuclear cross section describes the interactions of charged
leptons($l$) with nuclei via virtual photon exchange. The dominant part of
the cross section is from nearly real photons 
(negative four-momentum squared $Q^2\rightarrow 0$),
however, this is the limit where energy loss is negligible. 
Our approach here is to treat the $lN\rightarrow lX$ using the
deep-inelastic scattering formalism, and to use a nucleon structure
function $F_2$ consistent with data over the full range of $Q^2$.
In this approach, both soft physics at low $Q^2$ and hard perturbative
physics at high $Q^2$ are incorporated. The inclusion
of the perturbative physics at non-zero $Q^2$ has an effect on the
very high energy behavior of $\beta^{\rm nuc}$.

The standard variables used for $l (k)N(p)\rightarrow l (k')X$  
scattering
include
\begin{eqnarray}
q^2&=&(k-k')^2=-Q^2\\
x&=&{Q^2\over 2p\cdot q}\\
y&=&{p\cdot q\over p\cdot k}\ .
\end{eqnarray}
The differential cross section can be written in the form \cite{bk}
\begin{eqnarray}
{d\sigma(x,Q^2)\over dQ^2 dx}& =&
{4\pi\alpha^2\over Q^4}{F_2(x,Q^2)\over x}\Biggl[ 1-y-{Mxy\over 2 E}
\\ \nonumber
& &+ \Biggl(1-{2m_l^2\over Q^2}\Biggr){y^2(1+4M^2x^2/Q^2)\over 2(1+R(x,Q^2))}
\Biggr]\ .
\end{eqnarray}
The quantity $R$ is written in terms of $F_L$ and $F_1$ where
\begin{eqnarray}
R(x,Q^2)& = & {F_L(x,Q^2)\over 2xF_1(x,Q^2)}\\
F_L(x,Q^2)&=&\Biggl( 1+{4M^2x^2\over Q^2}\Biggr) F_2(x,Q^2)-2x F_1(x,Q^2)\ .
\end{eqnarray}
In all of the expressions above,
$m_l$ is the lepton mass and $M$ is the nucleon mass.
We have converted the differential cross section to a dependence on
$y$ and $Q^2$ and used the following limits of integration:
\begin{eqnarray}
Q_{\rm min}^2&\leq & Q^2\leq 2MEy -((M+m_\pi)^2-M^2)\\
y_{\rm min} &\leq & y\leq 1-{m_l/ E}\ .
\end{eqnarray}
where $Q_{\rm min}^2\simeq m_l^2y^2/(1-y)$ and $y_{\rm min}\simeq
((M+m_\pi)^2-M^2)/(2ME)$.
%We have used the fact that this is the inelastic formalism, so the
%minimum hadronic invariant mass is $M+m_\pi$.

The structure function $F_L(x,Q^2)$ is proportional to the longitudinal
photon-nucleon cross section.
In the $Q^2\rightarrow 0$ limit,
$F_L\sim Q^4$ while $F_1\rightarrow Q^2$, so $R\rightarrow 0$.
We have used $R(x,Q^2)$ modeled by Badelek, Kwieci\'nski
and Stasto in Ref. \cite{bks} for $10^{-7}<x<0.1$ and 0.01 GeV$^2<Q^2<50$
GeV$^2$. For $x>0.1$, the parameterization of Whitlow et al. \cite{whitlow}
is used. There is no evidence for target dependence of $R$.
The photonuclear $\beta^{\rm nuc}$ for muons 
evaluated with the $R$ parameterized by Badelek {\it et al.} differs 
only by a few percent from that calculated with $R=0$.
Consequently, in what follows, we set $R=0$. 

The nuclear structure function depends on the particular target.
The attenuation of quark density in a nucleus has been observed 
in deep inelastic lepton scattering (DIS) from nuclei 
at CERN \cite{cern} and Fermilab \cite{e665} energies 
in the region of small values of $x$ and 
$Q^2$.  
The data, taken over a
wide kinematic range  $10^{-5}<x<0.1$ and 0.05 ${\rm GeV}^2 <
Q^2< 100$ GeV$^2$, show a systematic reduction of nuclear structure function
$F_2^A(x,Q^2)/A$ with respect to the free nucleon structure function
$F_2^N(x,Q^2)$.  We define the shadowing ratio by
\begin{equation}
a(A,x,Q^2)={F_2^A(x,Q^2)\over A F_2^N(x,Q^2)}\ .
\end{equation}

The origin of nuclear shadowing effect is coherent 
scattering of the virtual photon off the nucleons inside the nucleus in 
the small $x$ region.  
The coherent multiple scattering
can be most conveniently handled by the 
Glauber diffractive approximation model \cite{glauber}.
The interactions of the virtual photons with nucleons is modeled differently
for high $Q^2$ and low $Q^2$. At low $Q^2$ ($< 1$ GeV$^2$),
calculations of shadowing \cite{lu} have often used the 
vector-meson-dominance (VMD) model \cite{vmd}, in which 
the virtual photon interacts with the nucleons via its hadronic fluctuations,
namely the $\rho$, $\omega$ and $\phi$ mesons. At higher $Q^2$, the
picture is instead that
the virtual photon interacts with partonic components of the nucleons
via its quark-antiquark pair ($q\bar q$) color-singlet 
fluctuation \cite{ina}.
Given
the high nucleon and parton densities, the quarks and gluons that belong
to different nucleons in the nucleus will recombine and annihilate, leading to
the so-called recombination effect first suggested by Gribov, Levin and
Ryskin \cite{glr} and later proven by Mueller and Qiu \cite{mq}. 
The net effect of either mechanism is that $F_2^A$ is lower than one
would expect by naive superposition ($AF_2^N$)
and weakly dependent on $Q^2$ in the range of interest for
the photonuclear cross section.  
We take a $Q^2$ independent function of $x$ and
$A$ consistent with the Fermilab E665 data taken in the
kinematic range of $0.0001<x<0.56$ and $0.1<Q^2<80$ GeV$^2$ \cite{e665}:
\begin{equation}
a(A,x,Q^2)\simeq a(A,x)=\cases{A^{-0.1} &  $ x<0.0014$\, ,\cr
        A^{0.069\log_{10}x+0.097}& $ 0.0014<x<0.04$\, ,\cr
	1 & $0.04<x $ \ .\cr}
\end{equation}

The structure function $F_2^A$ is approximated by
\begin{eqnarray}
F_2^A & =& a(A,x){A\over 2}(F_2^p +  F_2^n) \\ \nonumber
&=& a(A,x){A}{1\over 2}(1+P(x))F_2^p  \ ,
\end{eqnarray}
assuming $Z=A/2$.
Here $P(x)=1-1.85x+2.45 x^2-2.35 x^3+x^4$ describes the ratio
$F_2^n/F_2^p$, parameterized by the BCDMS experiment \cite{bcdms}.

The quantity $F_2^p(x,Q^2)$ is extracted in a variety of experiments 
in a range of $0<Q^2<5000$ GeV$^2$ and $5\times 10^{-6}<x<1$, though
kinematic limits restrict the range of $Q^2$ and
$x$ in a given experiment. The differential cross section must be integrated
from $Q^2=0$, where the perturbative QCD description of $F_2$ is not valid,
to values of $Q^2$ where QCD is valid. Consequently, a parameterization
of $F_2^p$ consistent with all the data is most useful for our purposes.
The parameterization of $F_2^p$ used here is the one by Abramowicz,
Levin, Levy and Maor (ALLM) \cite{allm}.
The ALLM parameterization involves two terms: a pomeron
contribution and a reggeon contribution. Parameters are used to fit all data
available from the pre-HERA era as well as H1 and ZEUS data published 
through 1997. The specific form with parameters
is detailed in the Appendix B. 

Eq. (3.4) shows the alternative to
the BB formula for $d\sigma/dy$ which appears in Appendix A. 
As an initial comparison of the two approaches, we show in Fig. 1  the
cross section for real photon-nucleon scattering, as a function of incident
photon energy, 
\begin{equation}
\sigma(\gamma N)=\lim_{Q^2\rightarrow 0}{4\pi^2\alpha F_2^N\over
Q^2}
\end{equation}
indicated by the solid line, and the Bezrukov-Bugaev cross section
(dashed line). Photon-proton data collected in Ref. \cite{pdg} are also
shown.
Our cross section agrees with the BB parameterization
at energies below $E_\gamma\sim 10^4$ GeV, however the BB cross section
increases more quickly with energy.   At 
$E_\gamma=10^9$ GeV, our cross section is $0.40$ mb, while the BB 
cross section is $0.58$ mb.  

For the calculation of the lepton propagation in rock, it is useful to 
compare
the $\gamma A$ cross section 
for standard rock $(A=22)$.  Here we include the nuclear shadowing effects.  
Our results are shown in Fig. 2 with the solid line. The BB cross section
is the dashed line.  
We note that our results are in 
agreement with the BB parameterization for $\gamma A$, namely
\begin{equation}
\sigma(\gamma A)=A\sigma(\gamma N)[0.75G(x)+0.25]\ ,
\end{equation}
where $x$, $G(x)$ and $\sigma(\gamma N)$ appear in Eq. (A11), 
over a wide energy.
The largest deviation occurs at the lowest photon energy shown, where
the photo-nuclear contribution is least important for charged lepton 
energy loss.

\section{Muon Energy loss, Survival Probability, and Range}

The expected
average muon energy loss in traversing a material of depth $\Delta X$ 
is characterized by
$\langle dE/dX\rangle$, indicated in Eq. (1.1).
The results for the standard bremsstrahlung, pair production and 
photonuclear (BB) differential cross sections summarized by LKV, 
and our results of using the ALLM differential
cross section are shown in Fig. 3. Our result for 
$\beta^{\rm nuc}$
begins to diverge from the standard values at $E\sim 10^6$ GeV,
and is a factor of about 1.6 higher at $E=10^9$ GeV.
In terms of the total $\beta$, the total with ALLM contributions is
a factor of 
1.15 larger than with the BB photonuclear contribution at $E=10^9$ GeV.  

To explore the effect of the slightly larger value of $\beta^{\rm nuc}$
for muons,
we have evaluated the muon survival probability in standard rock.  
The survival probability 
$P(E,X)$ for a muon to survive to a depth $X$ given incident
energy $E$ incorporates
the effects of fluctuations due to radiation. 
In Fig. 4, we show our muon survival probabilities (solid lines)
for $E=10^3-10^9$ GeV,
in decades of energy, versus survival depth $X$ (in km.w.e.) for
standard rock ($A=22$ and $\rho=2.65$ g/cm$^3$). We have taken 
$E_{min}=1$ GeV in the Monte Carlo.
Using the LKV defaults in our Monte Carlo program yields 
the dashed lines, which agree with the Lipari-Stanev
result in Ref. \cite{ls}.

The two sets of survival probabilities translate to average muon ranges
with incident energy $E$ and final energy $E_{min}=1$ GeV.
The average
range $\langle R(E)\rangle$ is defined by
\begin{equation}
\langle R(E)\rangle =\int_0^\infty dX\, P(E,X)\ .
\end{equation}
The average ranges for our calculation are shown in Fig. 5
by the solid lines. The standard LKV ranges that we have calculated using
the same muon transport Monte Carlo simulation are shown with dashed
lines. At $E=10^9$ GeV, the two calculations differ by only 5\%.
The deviation from the standard calculation increases with energy.

\section{Tau energy loss, Survival Probability, and Range}

Essentially the same procedure for calculating muon energy loss can
be applied to the tau lepton, with the important modification that the
tau has a decay length considerably shorter than the muon decay length.
We have evaluated the pair production, bremsstrahlung and ionization
energy loss according to the formulas in Appendix A for tau leptons,
and we have evaluated the tau photonuclear
differential cross section using the ALLM
expression in Eq. (3.4). Fig. 6 shows the tau energy loss
contributions to $\beta$ for standard rock. In this figure, $\beta$ is
plotted on a logarithmic scale because the bremsstrahlung is very
suppressed relative to the other contributions, due to the much heavier
lepton mass. The photonuclear contribution to $\beta$ dominates above
$E\sim 10^5$ GeV.

In the absence of energy loss, the tau survival probability corresponding
to the curves in Fig. 4 are exponentials of the form:
\begin{equation}
P(E,X)= \exp \Biggl[ -{X\over \gamma c \tau\rho} \Biggr],
\end{equation}
where $\gamma=E/m_{\tau}$ is the  Lorentz gamma factor, 
$c\tau = 86.93\ \mu$m is 
the tau decay length 
and $\rho$ is the material density.
The average range with no energy loss, as defined by Eq. (4.1), is just 
\begin{equation}
\langle R(E)\rangle_{decay}=\gamma c\tau\rho \ .
\end{equation}

For incident energies between $10^3$ GeV and $10^9$ GeV, we have
made a Monte Carlo evaluation of tau energy loss
in water, rock and iron including the
electromagnetic energy loss mechanisms as described in the Appendix.

At low energies, the decay length of the tau is short and
energy loss is relatively unimportant, so the
survival probability is just the exponential in Eq. (5.1). The
survival probability is increased at fixed depth
as the energy of the tau increases.
The survival probability curves can be put on the same plot by using
$P(E,X)$ versus $X/E$
as shown in Fig. 7 for water. In Fig. 7, 
the decay distribution Eq. (5.1) is indicated by a dashed line.
A solid line overlays it, which is the probability distribution, 
as computed
by our Monte Carlo, with incoming tau energy
$E=10^3$ GeV. The lower solid line is the
survival probability for incident tau energy $E=10^9$ GeV.
The probabilities are evaluated with a minimum tau energy of 50 GeV.
By changing the minimum energy to 100 GeV, we see no change in the
probability distributions for incident tau energies $E=10^{3}-10^{9}$ GeV.

In Fig. 8, we compare the tau decay length with the range of tau with incident energy E and 
final energy $E_{min}=50$ GeV. 
The dashed line shows the tau decay length (Eq. (5.2)), while
the solid line shows the evaluation of Eq. (4.1) including electromagnetic
energy loss for propagation in water.
The deviation from
the simple gamma factor scaled decay length starts at about $10^8$ GeV
in water,
and by $10^9$ GeV, the average range is 35\%  lower.
The dot-dashed and dotted lines in Fig. 8 show the tau ranges in standard
rock and iron. For tau propagation in iron, the tau range is an order
of magnitude shorter than its decay length for an incident energy
of $10^9$ GeV.

Our results differ from the estimate of the range by Fargion in Ref. 
\cite{fargion}, mainly because of the inclusion of the photonuclear and pair
production processes in electromagnetic energy loss. The estimate in Ref.
\cite{fargion} relies on an approximate solution to Eq. (1.1), where
$\alpha$ and $\beta$ are assumed to be energy independent, and $\beta$
is approximated numerically by rescaling the muon pair production $\beta$.
The resulting $\beta$ is more than an order of magnitude smaller than the
$\beta$ values we obtain, for example, in Fig. 6 for rock. As a consequence,
whereas in Ref. \cite{fargion}, electromagnetic energy loss is never relevant,
we have shown that accounting for photonuclear, bremsstrahlung and pair
production energy loss mechanisms reduces the tau range in water beginning at
$E\sim 10^8$ GeV, at even lower energies for more dense materials.

\section{DISCUSSION}

We have re-evaluated the muon energy loss due to photonuclear interactions
using the
recent HERA results for the real and virtual photon-nucleon scattering.
We have used the ALLM parameterization of the electromagnetic structure
function $F_2(x,Q^2)$
to evaluate the photonuclear cross section including the
$Q^2\gg0$ region.
As compared with the previous Bezrukov and Bugaev result, our approach yields
a change in $\beta$ for photonuclear interactions. 

In our evaluation of energy loss and effective ranges,
except for tau decays, we have ignored the effects of weak interactions
on the propagation of muons and taus. To set the scale for charged
lepton energy loss via weak interactions, 
$$\beta_{weak}\simeq N_A\langle y\rangle \sigma_{lN}(E)\ .$$
For $\langle y\rangle \simeq 0.2$ and using $\sigma_{lN}\sim
\sigma_{\nu N}$ from Ref. \cite{gqrs2}, we find that for 
$\beta_{weak}$ to be larger than a typical high energy electromagnetic
$\beta\simeq 10^{-6}$ cm$^2$/g, the charged lepton energy must be larger
than $\sim 10^{16}$ GeV. Below that energy, weak interactions do not play
a significant role in charged lepton energy loss.

Apart from energy loss, weak interactions also contribute to charged
lepton disappearance through charged-current interactions. The tau charged-%
current interaction length is comparable to its decay length at energies
above $10^{10}$ GeV in water. We have
shown in Fig. 8 that already at $10^8$ GeV, taus lose energy in water
due to electromagnetic interactions. For tau energies above $10^{10}$ GeV in
water,
it will be an interplay of electromagnetic energy loss, decay and
charged-current weak interaction disappearance of taus  that will dictate
tau effective ranges. Even in lead, which has a density of 11.35 g/cm$^3$,
the decay length is shorter than the charged-current interaction length 
for energies below a few times $10^9$ GeV, so our neglect of
weak interactions below $E=10^9$ GeV is a good approximation.

The results presented here are not significantly modified by the 
Landau-Pomeranchuk-Migdal (LPM) effect \cite{lpm}.
The LPM effect arises from low-momentum transfer lepton scattering, where
bremsstrahlung should, in principle, be evaluated including coherence effects
due to scattering in the medium. The effect has been measured in the case of
electron scattering \cite{lpmexpt} and summarized in Ref. \cite{klein}, 
but is suppressed for heavier leptons. The
suppression factor can be written crudely in terms of the photon energy
$q^0=y\cdot E$ as \cite{klein}
\begin{equation}
S(y,E)=\sqrt{{y\over 1-y}{E_{LPM}\over E}}
\end{equation}
when the argument of the square root is positive ($y<E/(E+E_{LPM})$),
where
\begin{equation}
E_{LPM}= 1.38\times 10^{13}{\rm GeV\over cm}\cdot X_0\ .
\end{equation}
The radiation length $X_0$ is 36.1 cm for water and 10.0 cm for rock.
For the lepton energies of interest here, $E<10^9$ GeV, this means that LPM
suppression for muons occurs for $y< 7\times 10^{-5}({\rm cm}/X_0)$.
In practice, this affects the continuous ``soft'' term in Eq. (2.1).
The approximation for $S(y,E)$ leads to a decrease in the soft
term by at most 0.1\%\ at the highest energies considered here.
This is consistent with, for example, the more detailed
evaluation of $\beta^{\rm brem}$
by Polityko {\it et al.} in Ref. \cite{polityko}, where deviations from
standard $\beta^{\rm brem}$
evaluations occur above $E=10^{11}$ GeV. 
The LPM effect is even further suppressed for taus, since $E_{LPM}$
scales like the lepton mass to the fourth power.

The change to $\beta$ from using the
ALLM photonuclear interaction cross section increases with
energy and reaches 60\%\ for $E=10^9$ GeV for muons. Still, for muons, the
photonuclear processes do not dominate $\beta$ for $E<10^9$ GeV, 
and the overall effect on $dE/dX$ is smaller.
The muon survival probabilities and
average muon ranges are obtained from the
one dimensional Monte Carlo approximation.  We find our results to be
within $5\%$ from the standard LKV value with
incident energy $E<10^9$ GeV and final energy $E_{min}=1$ GeV.

We have evaluated the tau energy loss including
the photonuclear processes as well as the ionization,
pair production and bremsstrahlung energy loss.
At energies above $\sim 10^8$ GeV, tau interactions
become important in water and the range becomes significantly shorter.
The effect appears at lower energies for more dense materials, for example,
at tau energies less than $10^7$ GeV for iron.
We thus expect a decrease in the observed tau flux and tau neutrino energy
relative to that expected if these effects had be ignored. The effect is
most important  for  high energy taus arriving from directions
just below the horizon. For directions subtending a significant portion
of the Earth, energies will be degraded to less than $10^6$ GeV by repeated
neutrino conversion to tau and tau decay. The last few steps in that process
will occur at energies where $dE/dX$ is not important. Near the horizon,
the interaction length for charged current conversion becomes comparable 
to the length of the chord through the Earth, and the produced
taus may arrive directly in the detector depending on their range. For
$E>10^8$ GeV, the tau range is decreased  in water and such taus will either 
not arrive or arrive at lower energies. Correspondingly, for taus
produced in a detector, measured $dE/dX$ would be higher than expected
without the revised photonuclear effects calculated here.

\acknowledgements
Work supported in part
by National Science Foundation Grant No.
PHY-9802403 and DOE Contract DE-FG02-95ER40906.
We thank the Aspen Center for Physics for its hospitality while this was
was initiated, and 
we thank A. Stasto for providing us with a computer program
to evaluate $R(x,Q^2)$. We acknowledge early conversations with R. Engel
about using $F_2$ in the photonuclear process.

\newpage

\appendix

\section{Energy Loss Formulas}

For completeness, we reproduce the standard formulas for ionization,
pair production and bremsstrahlung energy loss. 
These are also collected in the unpublished work of Lohmann, Kopp and
Voss \cite{lkv}.
The Bezrukov and Bugaev
(BB) parameterization of photonuclear energy loss is also included below.

The constants are defined as follows:
\begin{eqnarray}
\alpha & &  = 1/137\\ \nonumber
N_A  & & =6.023\times 10^{23},\ {\rm Avogadro's\ number}\\ \nonumber
Z  & & = 
{\rm \ atomic \ number},\  A={\rm atomic\ weight\ of\ medium} \\ \nonumber
m_e,  & & \ m_l {\rm\ rest\ masses\ of\ electron\ and\ muon \ or\ tau}
\\ \nonumber
\lambda_e  & & =3.8616\times 10^{-11}\  {\rm cm,
\ Compton\ wavelength\ of\ the\ electron}\\ \nonumber
e  & & = 2.718\\ \nonumber
\end{eqnarray}

To account for energy loss on atomic electrons, the factor $Z^2$ in the 
bremsstrahlung and pair production formulas below should be replaced by
$Z(Z+1)$.

\subsection{Ionization Energy Loss}

The ionization loss for a incident muon or tau of energy $E$
and momentum $p$ is
given by the Bethe-Bloch formula \cite{bloch}:

\begin{equation}
{dE\over dX}=\alpha^2 2\pi N_A\lambda_e^2 {Z m_e\over A\beta^2}
\Biggl( \ln
\frac{2m_e\beta^2\gamma^2E'_m}{I^2(Z)}-2\beta^2+\frac{E'{_m}{^2}}
{4E{^2}}-\delta(X)\Biggr)
\end{equation}
where
$$\beta = p/E\ ,$$
$$\gamma=E/m_l\ .$$
The quantity 
\begin{equation}
E_m^\prime = 2m_e{p^2\over m_e^2+m_l^2+2m_e E}
\end{equation}
is the maximum energy that can be transfered to the electron.
A density correction is parameterized by $\delta(X)$ for 
$X= \log_{10}(\beta\gamma)$ by 
Sternheimer {\it et al.} \cite{sternheimer} as
\begin{eqnarray}
\delta (X) & &  = 4.6052X + a (X_1-X)^m + C \hspace{1in} X_0<X<X_1\\ \nonumber
\delta (X) & &  = 4.6052X + C \hspace{2.1in} X>X_1\ .\\ \nonumber
\end{eqnarray}
The values of $X_0,\ X_1,\ a,\ m,\ C$ and the
mean ionization potential $I(Z)$, 
for rock and water  are given in Table 2.

\subsection{Bremsstrahlung Energy Loss}

Following Petrukhin and Shestakov's evaluation of the form factors in
the Bethe-Heitler bremsstrahlung differential cross section \cite{ps}, the
differential cross section used here is:
\begin{equation}
{d\sigma\over dy}=\alpha^3\Biggl( 2Z\lambda_e {m_e\over m_l}\Biggr)^2{1\over y}
\Biggl({4\over 3}-{4\over 3}y+y^2\Biggr) \phi(\delta)\ ,
\end{equation}
where
\begin{eqnarray}
\phi(\delta) & & = \ln \Biggl[ { {189 m_l\over m_e}Z^{-1/3} \over
1+{189\sqrt{e}\over m_e}\delta Z^{-1/3}} \Biggr], \ \quad Z\leq 10\\ \nonumber
\phi(\delta) & & = \ln \Biggl[ { {2\over 3}{189 m_l\over m_e}Z^{-2/3} \over
1+{189\sqrt{e}\over m_e}\delta Z^{-1/3}} \Biggr], \ \quad Z> 10\\ \nonumber
\delta & & = {m_l^2 y\over 2 E (1-y)}\ .
\end{eqnarray}
The range of $y$ integration is
\begin{equation}
0\leq y\leq 1-{3 m_l\over 4 E}\sqrt{e}Z^{1/3}\ .
\end{equation}

\subsection{Pair Production Energy Loss}

Pair production energy loss depends on the asymmetry parameter $\rho$ given
as
\begin{equation}
\rho = {E^+-E^-\over E^++E^-}, 0\leq |\rho| \leq \Biggl(1-{6m_l^2\over
E^2(1-y)}\Biggr)\sqrt{1-{4m_e\over E y}}
\end{equation}
given $E^+$ and $E^-$, the energies of the positron and electron pair.
The differential cross section is parameterized by \cite{kp}
\begin{equation}
{d^2\sigma\over dy d\rho} = \alpha^4{2\over 3\pi} (Z\lambda_e)^2{1-y\over y}
\Biggl( \phi_e+{m_e^2\over m_l^2}\phi_l \Biggr)\ .
\end{equation}
where
the functions $\phi_e$ and $\phi_l$ are as follows:
\begin{eqnarray}
\phi_e& & = \Biggl[\Biggl( (2+\rho^2)(1+\beta) +
\xi(3+\rho^2)\Biggl)\ln(1+{1\over\xi}
) + {{1-\rho^2-\beta}\over{1+\xi}} - (3+\rho^2)\Biggr]L_e
\\ \nonumber
\phi_l& & = \Biggl[\Biggl( (1+\rho^2) (1+{3\over2}\beta) -{1\over\xi}
(1+2\beta) 
(1-\rho^2) \Biggl) \ln(1+\xi) \\\nonumber
& & + {{\xi(1-\rho^2-\beta)}\over{1+\xi}} + (1+2\beta)(1-\rho^2)\Biggr]L_l
\\ \nonumber
L_e& & =\ln {RZ^{-{1\over3}}\sqrt{(1+\xi)(1+Y_e)}\over1+
{{2m_e\sqrt eRZ^{-{1\over3}}(1+\xi)(1+Y_e)}\over {Ey(1-\rho^2)}}}
-{1\over2}\ln
\Biggl[1+({3\over2}{m_e\over m_l}Z^{1\over3})^2(1+\xi)(1+Y_e)\Biggl]
\\ \nonumber
L_l & & =\ln {RZ^{-{2\over3}}{2\over3}{m_l\over m_e}\over1+
{{2m_e\sqrt eRZ^{-{1\over3}}(1+\xi)(1+Y_l)}\over {Ey(1-\rho^2)}}}
\\ \nonumber
Y_e& & ={ 5-\rho^2+4\beta(1+\rho^2)\over
2(1+3\beta)\ln(3+{1\over\xi})-\rho^2-2\beta(2-\rho^2)}
\\ \nonumber
Y_l & & ={ 4+\rho^2+3\beta(1+\rho^2)\over
(1+\rho^2)({3\over2}+2\beta)\ln(3+\xi)+1-{3\over2}\rho^2}
\\ \nonumber
\beta& & ={y^2\over2(1-y)},\quad \xi=\Biggl({m_\l y\over 2m_e}\Biggr)^2
{(1-\rho^2)\over(1-y)}\ .
\end{eqnarray}
The value of $R$ is $R=189$.
The range of $y$ integration is
\begin{equation}
{4m_e\over E}\leq y\leq 1-{3 m_l\over 4 E}\sqrt{e}Z^{1/3}\ .
\end{equation}

\subsection{Bezrukov and Bugaev Photonuclear Energy Loss}

The parameterization of the Bezrukov and Bugaev photonuclear energy
loss is \cite{bb}
\begin{eqnarray}
{d\sigma\over dy} = {\alpha \over 2\pi} A \sigma_{\gamma N}y
\Biggl[{3\over4}G(x)\Biggl(\kappa\ln(1+{m_1\over t})-
{\kappa m_1^2\over m_1^2+t}-{2m_l^2\over t}\Biggr) + \\ \nonumber
{1\over4}\Biggl(\kappa\ln(1+{m_2\over t})-{2m_l^2\over t}
\Biggr) + {m_l^2\over 2t}\Biggl({3\over4}G(x) {m_1^2\over m_1^2+t}+
{1\over4}{m_2^2\over t}\ln(1+{t\over m_2^2})\Biggr)\Biggr]
\end{eqnarray}
where
\begin{eqnarray}
G(x)&&={3\over x^3}\Biggl({x^2\over2}-1+e^{-x}(1+x)\Biggl)\\\nonumber
x&&=0.00282A^{1\over3}(\sigma_{\gamma N}(E/{\rm GeV})/\mu{\rm b})\\\nonumber
\sigma_{\gamma N}(E)&&=114.3 + 1.647
\ln^2(0.0213\,E/{\rm GeV})\ \mu {\rm b}\\\nonumber
t&&= {m_l^2y^2\over (1-y)}\\\nonumber
\kappa&&=1-{2\over y} +{2\over y^2}\\\nonumber
m_1^2&&=0.54 \ {\rm GeV}^2\\\nonumber
m_2^2&&=1.8\ {\rm GeV}^2\nonumber
\end{eqnarray}

\section{ALLM Parameterization of $F_2$}
The ALLM phenomenological 
parameterization of $F_2^p(x,Q^2)$ has a pomeron and reggeon
contribution \cite{allm}:
\begin{equation}
F_2^p(x,Q^2)={Q^2\over Q^2+m_0^2}\Bigl(
F_2^{\cal P}(x,Q^2)+F_2^{\cal R}(x,Q^2)\Bigr)\ ,
\end{equation}
where the two terms have the form
\begin{eqnarray}
F_2^{\cal P}(x,Q^2)&=& c_{\cal P}(t)x_{\cal P}^{a_{\cal P}(t)}(1-x)
^{b_{\cal P}(t)} \\
F_2^{\cal R}(x,Q^2)&=& c_{\cal R}(t)x_{\cal R}^{a_{\cal R}(t)}(1-x)
^{b_{\cal R}(t)}\ . 
\end{eqnarray}
The functions $c_{\cal P}(t)$, etc., are parameterized in terms of
\begin{equation}
t  =  \ln\Biggl( {\ln{(Q^2+Q_0^2)/\Lambda^2}\over \ln{Q_0^2/\Lambda^2}
}\Biggr)\ .
\end{equation}
The form for $f=c_{\cal R}$ and $a_{\cal R}$ is
\begin{equation}
f =  f_1+f_2t^{f_3}\ ,
\end{equation}
for $g=c_{\cal P}$ and $a_{\cal P}$,
\begin{equation}
g =  g_1+(g_1-g_2)\Biggl[{1\over 1+t^{g_3}}-1\Biggr]\ ,
\end{equation}
and finally, $h=b_{\cal R}$ and $b_{\cal P}$,
\begin{equation}
h =  h_1^2+h_2^2t^{h_3}\ .
\end{equation}
In the expressions for $F_2^{\cal P}$ and $F_2^{\cal R}$,
\begin{eqnarray}
x_{\cal P} & = & {Q^2+m_{\cal P}^2\over Q^2+m_{\cal P}^2+W^2-M^2}\\
x_{\cal R} & = & {Q^2+m_{\cal R}^2\over Q^2+m_{\cal R}^2+W^2-M^2} \ .
\end{eqnarray}
The parameters satisfying the fits described in Ref. \cite{allm}
are shown in Table II.

\newpage
\begin{table}
\caption{Parameter values in ionization energy loss [12].} 
\begin{tabular}{lccccccccc}
Material & $I$/eV & -C & $X_0$ & $X_1$ & $a$ & $m$ & $Z$ & $A$ & 
$\rho$/(g/cm$^3$) \\
Water & 75.0 
      &  3.502 
      &  0.240 
      &  2.800 
      &  0.091 
      &  3.477 
      &  6.6 
      &  11.89 
      &  1.00 \\
Standard Rock & 136.4
      & 3.774 
      & 0.049 
      &  3.055
      &  0.083
      &  3.412
      &  11 
      &  22 
      &  2.65 \\
Iron & 286.0
      & 4.291 
      & -0.0012 
      &  3.153
      &  0.147
      &  2.963
      &  26 
      &  55.84 
      &  7.87 \\
\end{tabular}
\end{table}

\begin{table}
\caption{Parameter values for ALLM parameterization of $F_2^p(x,Q^2)$.}
\begin{tabular}{lccc}
$c_{{\cal P}1}$ & 0.28067 & $c_{{\cal R}1}$ & 0.80107 \\
$c_{{\cal P}2}$ & 0.22291 & $c_{{\cal R}2}$ & 0.97307 \\
$c_{{\cal P}3}$ & 2.1979 & $c_{{\cal R}3}$ & 3.4942 \\
$a_{{\cal P}1}$ & -0.0808 & $a_{{\cal R}1}$ & 0.58400 \\
$a_{{\cal P}2}$ & -0.44812 & $a_{{\cal R}2}$ & 0.37888 \\
$a_{{\cal P}3}$ & 1.1709 & $a_{{\cal R}3}$ & 2.6063 \\
$b_{{\cal P}1}$ & 0.60243 & $b_{{\cal R}1}$ & 0.10711 \\
$b_{{\cal P}2}$ & 1.3754 & $b_{{\cal R}2}$ & 1.9386 \\
$b_{{\cal P}3}$ & 1.8439 & $b_{{\cal R}3}$ & 0.49338 \\
$m_0^2$(GeV$^2$) & 0.31985 & $m_{\cal P}^2$(GeV$^2$) & 49.457 \\
$m_{\cal R}^2$(GeV$^2$) & 0.15052 & $Q_0^2$(GeV$^2$) & 0.46017 \\
$\Lambda^2$(GeV$^2$) & 0.06527 & & \\
\end{tabular}
\end{table}
%\end{appendix}

\vfil\eject

\newpage
%figure 1

\begin{figure}
\centerline{\psfig{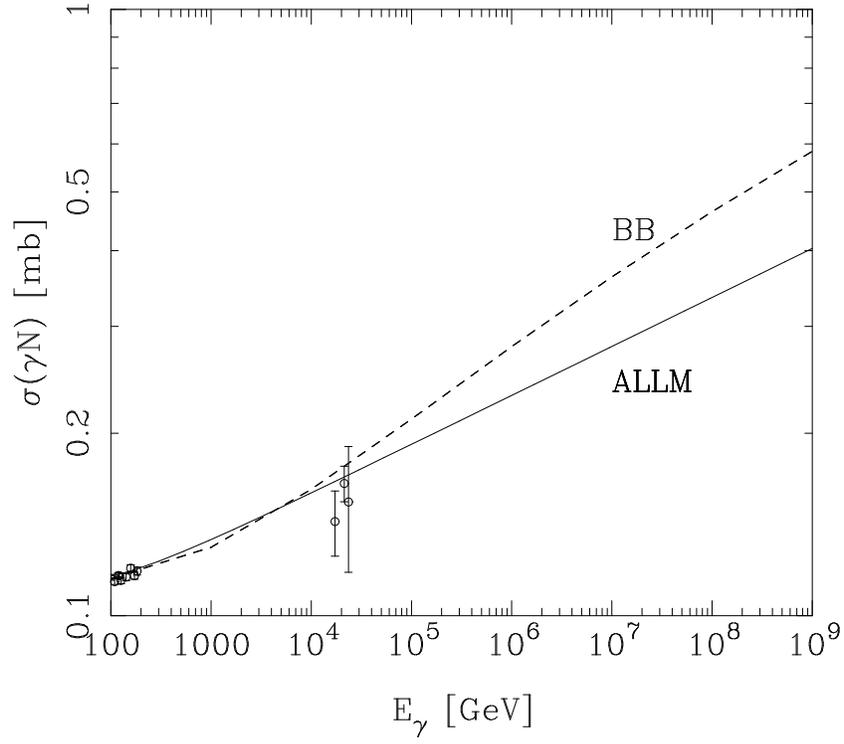}}
\vspace{0.3in}
\caption
{The photon-nucleon cross section as a function of incident 
photon energy for the BB (dashed) and ALLM (solid) parameterizations. Also shown are 
photon-proton data collected in Ref. [30].
}
\end{figure}
\vfil\break

%figure 2
\begin{figure}
\centerline{\psfig{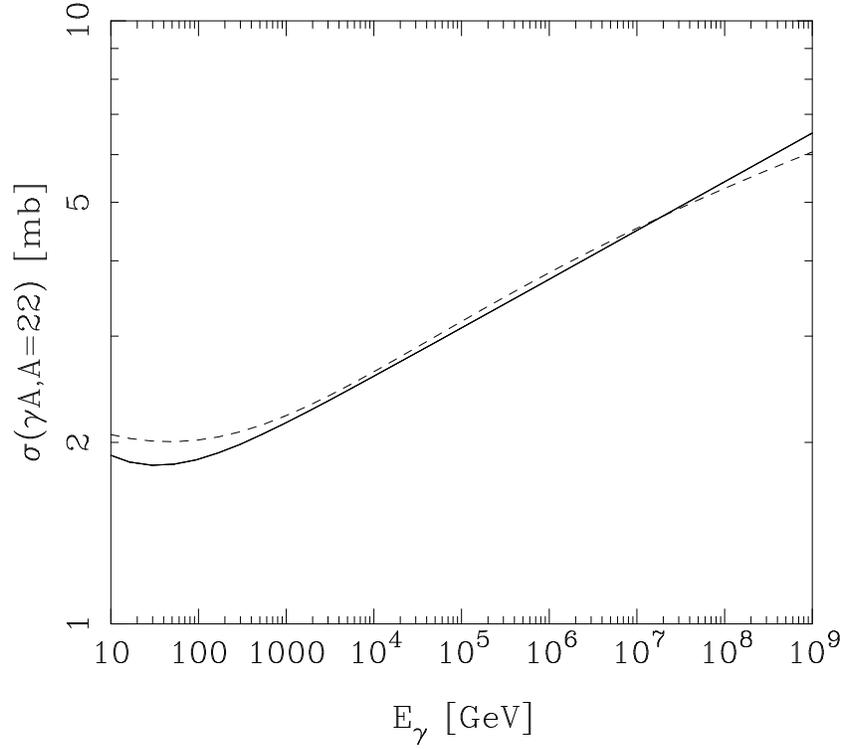}}
\vspace{0.3in}
\caption{The photon-nucleus cross section 
standard rock ($A=22$), as a function
of incident photon energy $E$
using the ALLM parameterization and Eqs. (3.9) and (3.10) 
for the shadow factor
and conversion to nucleon structure function. 
}
\end{figure}
\vfil\break

%figure 3
\begin{figure}
\centerline{\psfig{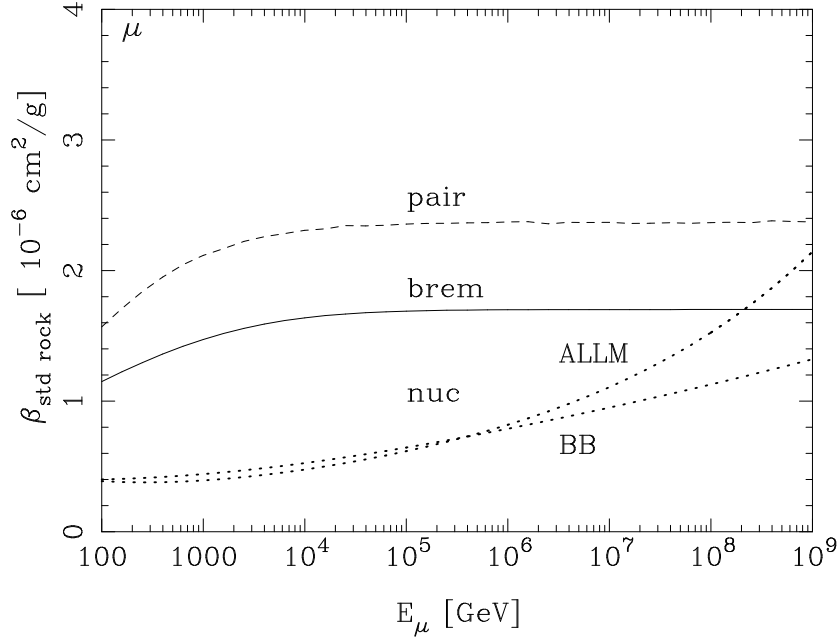}}
\vspace{0.3in}
\caption{The $\beta$ value for muon in standard rock ($A=22$), including bremsstrahlung
(solid line), pair production (dashed) and photonuclear (dotted) interactions.
}
\end{figure}
\vfil\break

%figure 4
\begin{figure}
\centerline{\psfig{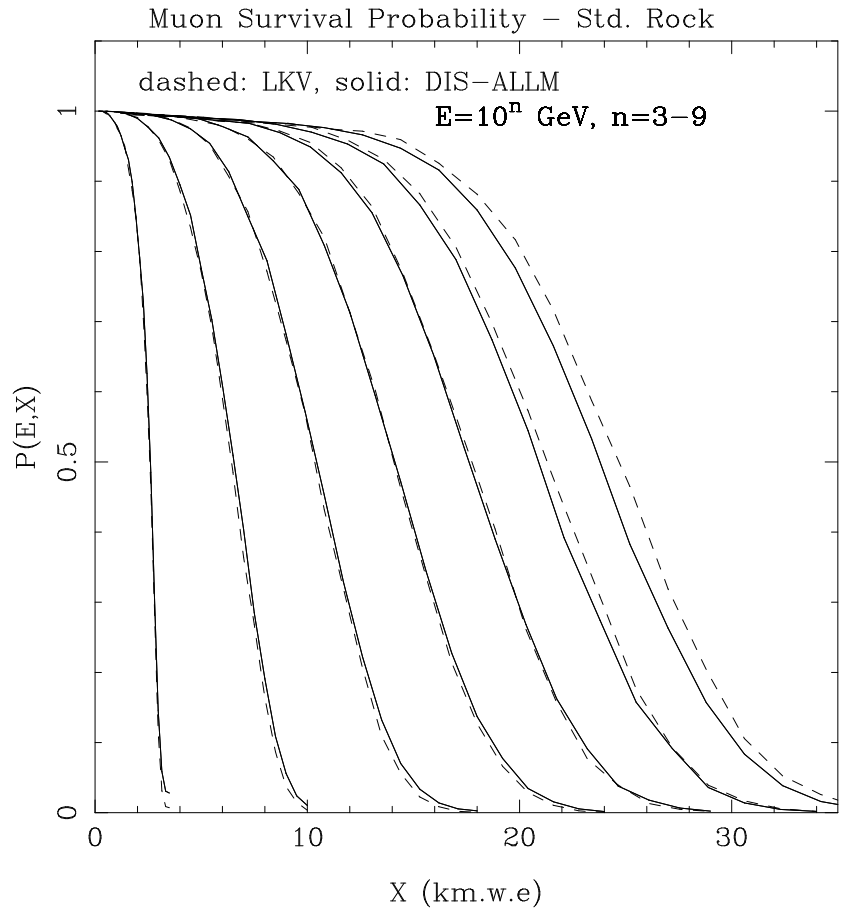}}
\vspace{0.3in}
\caption
{Muon survival probabilities in rock
using BB differential cross section
for the photonuclear term (dashed) and the ALLM differential cross section
(solid).
}
\end{figure}
\vfil\break

%figure 5
\begin{figure}
\centerline{\psfig{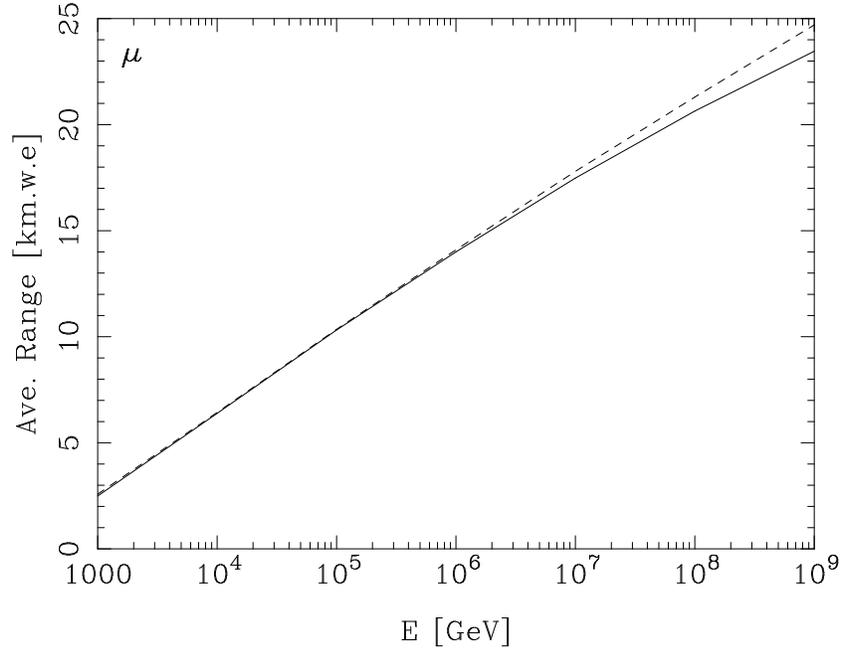}}
\vspace{0.3in}
\caption{Average muon range in standard rock (km.w.e. depth), for
incident muon energy $E$, final muon energy $E_f\geq 1$ GeV, using
the standard LKV treatment of energy loss, including the BB differential
cross section (dashed) and substituting the ALLM photonuclear calculation
(solid).
}
\end{figure}
\vfil\break

%figure 6
\begin{figure}
\centerline{\psfig{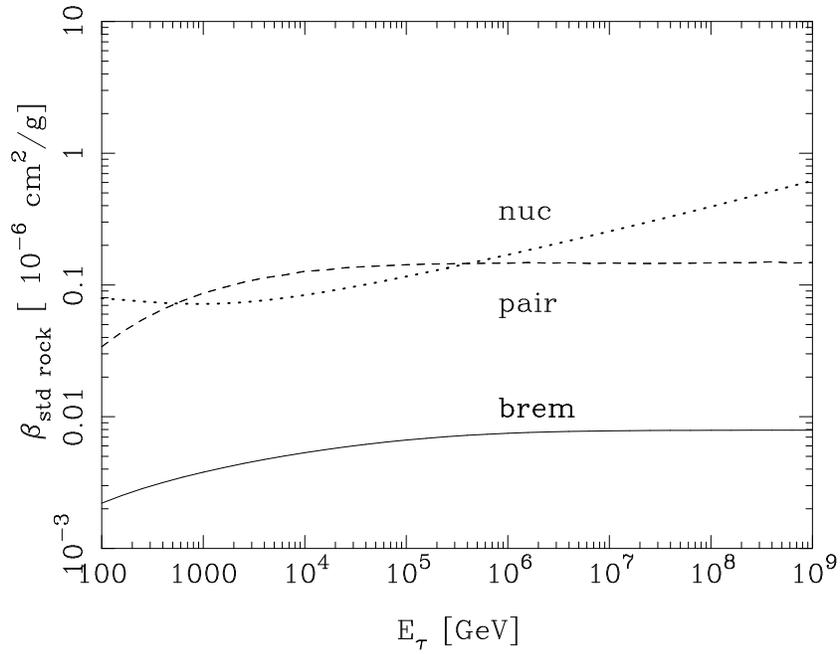}}
\vspace{0.3in}
\caption{The $\beta$ value for tau
in standard rock ($A=22$), including bremsstrahlung
(solid line), pair production (dashed) and photonuclear
(ALLM) (dotted) interactions.
}
\end{figure}
\vfil\break

%figure 7
\begin{figure}
\centerline{\psfig{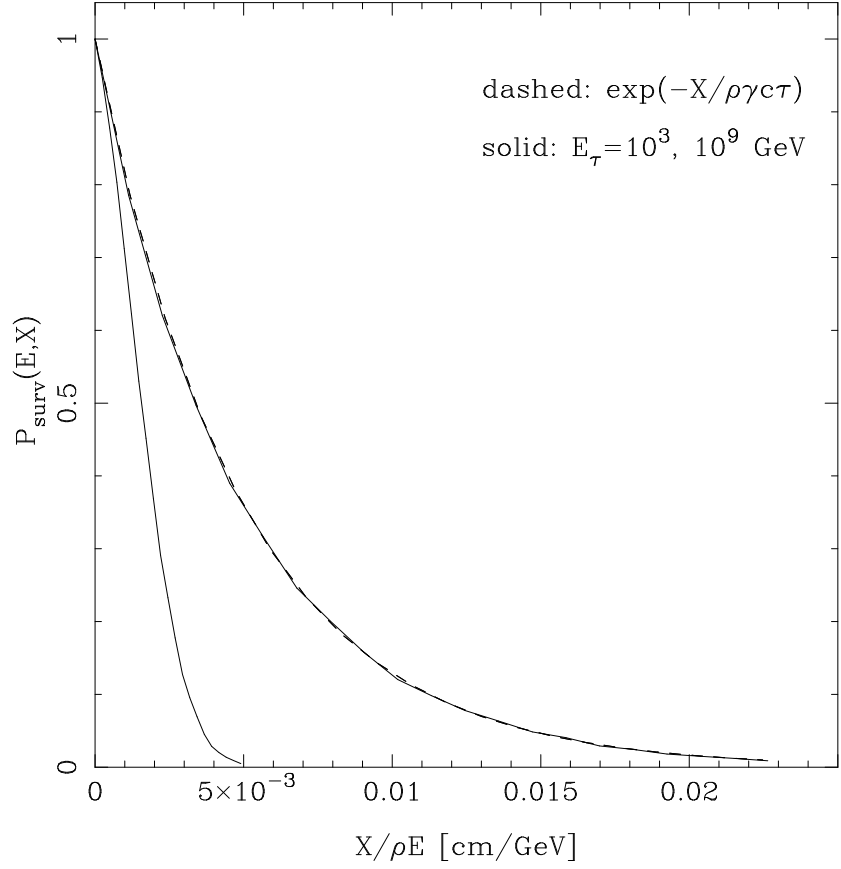}}
\vspace{0.3in}
\caption{
Decay distribution(dashed line) and survival probability curve for 
incoming tau energy
$E=10^3$ GeV (upper solid line) and $E=10^9$ GeV (lower solid line)
in water ($\rho=1$ g/cm$^3$).
}
\end{figure}
\vfil\break

%figure 8
\begin{figure}
\centerline{\psfig{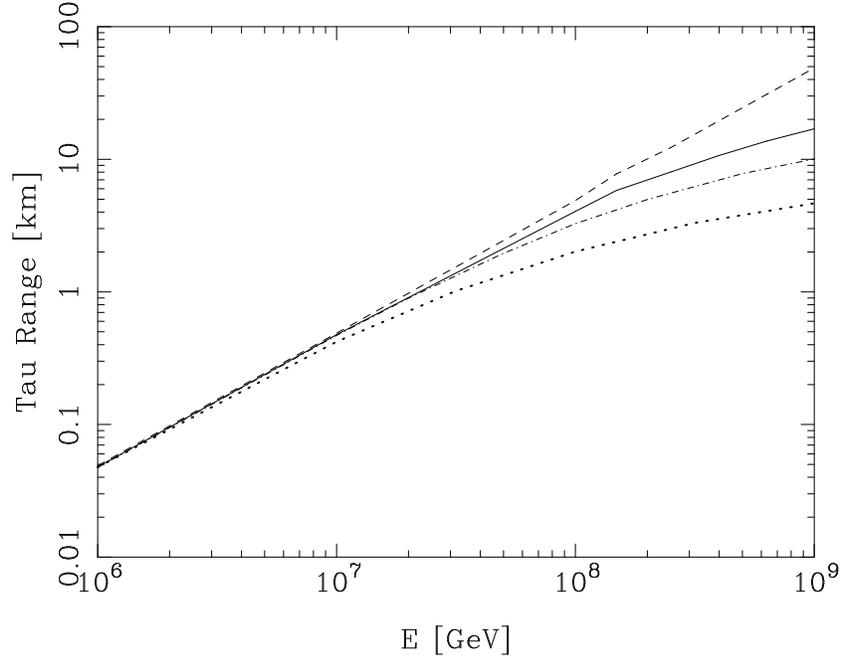}}
\vspace{0.3in}
\caption{Tau decay length (dashed line) and the 
average tau range in water (solid line), in rock (dot-dashed line)
and in iron (dotted line), for
incident tau energy $E$, final tau energy larger
than $E_{min}= 50$ GeV including electromagnetic energy loss.
}

\end{figure}
\end{document}